\documentclass[preprint,pdftex]{aastex}

\usepackage{mathtools}
\usepackage{upgreek}
\usepackage{graphicx}

\newcommand{\true}{\text{true}}
\newcommand{\fit}{\text{fit}}
\newcommand{\basis}{\text{best}}
\newcommand{\rms}{\text{RMS}}
\newcommand{\dd}{\text{d}}

\shorttitle{Designing Imaging Surveys for Retrospective Relative Photometric Calibration}
\shortauthors{Holmes et al.}

\begin{document}

\title{Designing Imaging Surveys for a Retrospective Relative Photometric Calibration}


\author{Rory Holmes, David W. Hogg\altaffilmark{1} and Hans-Walter Rix}
\affil{Max-Planck-Institut f\"ur Astronomie, K\"onigstuhl 17, Heidelberg, 69117, Germany.}
\altaffiltext{1}{Center for Cosmology and Particle Physics, Department of Physics, New York University, 4 Washington Place, New York, NY 10003, USA.}

\begin{abstract}
In this paper, we investigate the impact of survey strategy on the performance of self-calibration when the goal is to produce accurate photometric catalogs from wide-field imaging surveys. This self-calibration technique utilizes multiple measurements of sources at different focal-plane positions to constrain instruments' large-scale response (flat-field) from survey science data alone. We create an artificial sky of sources and synthetically observe it under four basic survey strategies, creating an end-to-end simulation of an imaging survey for each. These catalog-level simulations include realistic measurement uncertainties
and a complex focal-plane dependence of the instrument response. In the self-calibration step, we simultaneously fit for all the star fluxes and the parameters of a position-dependent flat-field.  For realism, we deliberately fit with a wrong noise model and a flat-field functional basis that does not include the model that generated the synthetic data.  We demonstrate that with a favorable survey strategy, a complex instrument response can be precisely self-calibrated. We show that returning the same sources to very
different focal-plane positions is the key property of any survey strategy designed for accurate retrospective calibration of this type. The results of this work suggest the following advice for those considering the design of large-scale imaging surveys: Do not use a regular, repeated tiling of the sky; instead return the same sources to very different focal-plane positions.
\end{abstract}

\keywords{Catalogs --- methods: data analysis --- techniques: photometric}

\section{Introduction}
Astronomers tend to think conventionally in terms of taking science data and calibration data separately. The former is used for scientific measurements of astronomical sources and the latter is used to constrain instrument parameters, such as the instrument response, the dark currents and so-on. But typically far more photons, or readout electrons, are collected during science exposures themselves; are these not incredibly constraining on the calibration? Indeed, in the retrospective photometric calibration of the Sloan Digital Sky Survey imaging data (SDSS), much more calibration information was obtained from the science data than the calibration data \citep{pad08}. But, of course, the SDSS imaging strategy had to be adjusted to make this calibration work: good redundancy was required in the data stream, and a redundancy of a very specific kind. Self-calibration has been used to obtain the most precise photometric calibration of the PanSTARRS imaging data \citep{sch12}.

In this paper, we argue that the next generation of large-scale imaging surveys should have their observation strategies designed from the very start with this kind of ``self-calibration'' in mind. This work focuses on the relative photometric calibration of a typical imaging instrument only, although similar techniques could be used to constrain many other calibration parameters, such as the optical distortion, point-spread-functions and so-on. The self-calibration technique utilizes the multiple measurements of sources in the survey arising from overlapping pointings. If these redundant measurements are recorded at different focal-plane positions or at different times within the survey, it is possible to constrain the relative instrument response by requiring that the post-calibration measurements yield consistent flux values. Through end-to-end, catalog-level survey simulations, we aim to identify the important properties of survey strategies that makes them advantageous for this kind of self-calibration. 

We construct realistic survey catalogs through mock observations of a synthetic sky according to a specified survey strategy. These simulations include a complex, position-dependent instrument response for the imaging instrument. Through the self-calibration procedure, we recover this instrument response by fitting a model that best describes the survey catalog. By comparing the \textit{fitted} instrument response to the \textit{true} instrument response we are able to assess the performance of the self-calibration procedure with different survey strategies. 

In this work we do not produce pixelated images.  Instead we create catalog-level simulations with realistic measurement uncertainties, including unmodeled source variability.  Complex effects are included in the simulations that are (deliberately) not correctly modeled at the analysis stage, in order to simulate the effects of unknown systematic errors within the catalog.

In Section \ref{sec:survey_simulations} we introduce the simulation chain constructed to produce the realistic survey catalogs. Section \ref{sec:self_cal} goes onto to detail the self-calibration procedure, with Section \ref{sec:metrics} summarizing the metrics we use to assess its performance. In Section \ref{sec:simple_surveys} we focus on four simple survey strategies, which allow us to draw conclusions on the performance of the self-calibration procedure with different survey properties. 

\section{Survey Catalog Simulations}
\label{sec:survey_simulations}
We have constructed an end-to-end simulation chain that produces a realistic imaging catalog from a specified survey strategy\footnote{All code used in this work is publicly available at \url{http://github.com/davidwhogg/SelfCalibration/}}. In this work we have kept the simulation parameters intentionally flexible, so that the sensitivity of the self-calibration procedure to different values can be investigated (the fiducial values are given in Table \ref{tab:parameters}). The simulations are split into a number of steps. The first is the generation of a synthetic sky. With a given pointing, single mock observations are performed on this sky. To build up a full-survey catalog, multiple single observations are performed according to the specified survey strategy. In this section, we detail the assumptions and methods used in each of these steps. 

\subsection{Synthetic Sky}
We generate a representative synthetic sky based on realistic object densities in the AB magnitude range $m_\text{min} = 17$ to $m_\text{max} = 22$~mag$_\text{AB}$, with these limits chosen to be consistent with the saturation and a $10\upsigma{}$ limits of deep, space-based, near-infrared, imaging survey. Sources are generated with random coordinates (uniformly distributed within the sky region being investigated) and with random magnitudes $m$ distributed according to
\begin{eqnarray*}
\log_{10} \frac{\dd N}{\dd m \, \dd \Omega} = a + b\,m + c\,m^2 \label{eqn:power_law} \quad , 
\end{eqnarray*}
where $\dfrac{\dd N}{\dd m \, \dd  \Omega}$  is the density of sources $N$ per unit magnitude $m$ and per unit solid angle $\Omega$, and $a$, $b$ and $c$ are model parameters. Even though our simulations make no distinction between galaxies and stars, the values of the parameters are found from fitting the Y-band galaxy populations reported in \citet{win11} only. These parameters were found to be $a = -13.05$, $b = 1.25$ and $c = -0.02$. The source magnitudes $m$ are related to the source fluxes $s$ simply by: $m = 22.5 - 2.5\log_{10}(s)$, where the 22.5 puts the fluxes in units of nanomaggies (nmgy). To limit the data, we only select the brightest sources within the survey area, up to a source density $d=300$~deg$^{-2}$, for the self-calibration procedure.

\subsection{A Single Exposure}
\label{sec:single_exposure}
With a telescope pointing $(\alpha, \beta)$ and camera orientation $\theta$, we find the focal-plane positions of the sources on the synthetic sky that fall within the instrument's field-of-view. In our simulations, we use a large instrument field-of-view of $0.75~ \text{deg} \times 0.75~\text{deg}$; a size consistent with current large survey imagers. An example of a single pointing exposure is shown in Figure \ref{fig:single_image}.

\subsubsection{Measured Count Rates}
To generate photometric catalogs, the true source fluxes $s_\true$ are converted into measured count rates $c$ with an complex, position-dependent instrument response model $f_\true$ and a measurement noise model. For a measurement $i$, the count rate $c_i$ recorded from a source $k$ depends on the \textit{true} instrument response $f_{\true}(\vec{x_i} | \vec{q}_\true)$---which is a function of focal-plane position $\vec{x_i}$---and the source's true flux $s_{k,\true}$ ,
\begin{eqnarray*}
c_i = f_{\true}(\vec{x_i} | \vec{q}_\true) \, s_{k, \true} + e_{i} \quad ,
\end{eqnarray*}
where $\vec{q}_\true$ are the parameters defining the \textit{true} instrument response (see Section \ref{sec:instrument_response_model}), and $e_i$ is a noise contribution drawn from the Normal Distribution $N(e|0,{\sigma_\true}^2)$. 

\subsubsection{Noise Model}
\label{sec:noise}
To construct the noise model, the simulated exposures are assumed to be background limited and that, for systematic reasons, there is an upper limit on the signal-to-noise ratio of 500 for bright sources. The noise model is complicated further by applying an extra term $\epsilon_i$ to the count rates' uncertainty variance, which we intentionally do not take into account in the analysis in order to simulate systematic problems with the instrument noise model. The \textit{true} noise model is therefore
\begin{eqnarray}
\sigma_{i, \true}^{2} = (1 + \epsilon_i) \, \delta^{2} + \eta^{2}\, [ f_\true(\vec{x_i} | \vec{q}_\true) \, s_{k, \true} ]^2 \quad , \label{eqn:noise}
\end{eqnarray}
where $\delta$ and $\eta$ are both constants and $\epsilon_i$ is a random number, drawn uniformly in the range [0.0, $\epsilon_{max}$), generated for each measurement $i$. The $m = 22$ mag 10$\upsigma$ detection limit introduced previously and the limit S/N$<500$ are used to set $\delta = 0.1585$ and $\eta = 0.0017$. The $\epsilon_i$ contribution is not taken into account in the analysis and therefore the uncertainty variances on the count rates are \textit{assumed} (incorrectly, but realistically) to be
\begin{eqnarray*}
\sigma_{{i}}^{2} = \delta^{2} + \eta^{2} \, c^{2}_i \quad 
\end{eqnarray*}
during this stage. 

\subsubsection{True Instrument Response Model}
\label{sec:instrument_response_model}
\begin{figure}
\includegraphics[width=\textwidth]{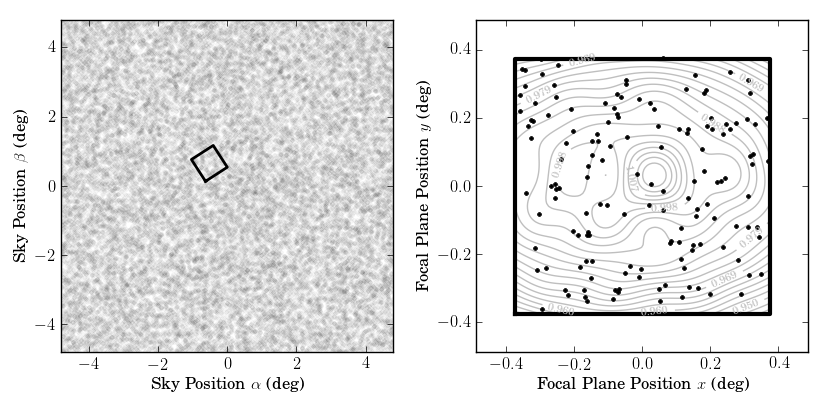}
\figcaption{A single exposure of the synthetic sky. Left: A plot of the bright sources within the synthetic sky used in the self-calibration procedure, with the focal-plane footprint overlaid. Right: The resultant distribution of the sources on the instrument's focal plane. The \textit{true} instrument model $f_{\true}(\vec{x_i} | \vec{q}_\true)$ is shown with contours.\label{fig:single_image}}
\end{figure}

We construct a complex, position independent instrument response model $f_\true(\vec{x_i} | \vec{q}_{\true})$ from a superposition of large- and small-scale variations:
\begin{eqnarray*}
f_\true(\vec{x_i} | \vec{q}_{\true, 1 \ldots 260}) = f_\text{large}(\vec{x_i} | \vec{q}_{\true, 1 \ldots 6}) + f_\text{small}(\vec{x_i} | \vec{q}_{\true, 7 \ldots 260}) \quad ,
\end{eqnarray*}
where $\vec{x_i} = (x_i, y_i)$ is the focal-plane position that the $k$th source falls at during the $i\text{th}$ measurement and $\vec{q}_\true$ are the parameters defining the instrument response model. This response model subsumes various effects commonly discussed in terms of detector flat-fielding and optical vignetting. The large-scale instrument response $f_\text{large}(\vec{x_i} | \vec{q}_{\true, 1 \ldots 6})$ is modeled as a second order polynomial:
\begin{eqnarray*}
f_\text{large}(\vec{x_i} | \vec{q}_{\true, 1 \ldots 6}) = q_{\true, 1} + q_{\true, 2} \, x_i + q_{\true, 3} \, y_i + q_{\true, 4} \, x_i^2 + q_{\true, 5} \, x_i \, y_i  + q_{\true, 6} \, y_i^2  \quad .
\end{eqnarray*}
The small-scale instrument response, which is constructed from sine and cosine contributions, is superimposed on this large-scale instrument response. The small-scale instrument response $f_\text{small}(\vec{x_i} | \vec{q}_{\true, 7 \ldots 260})$ is modeled as
\begin{eqnarray*}
f_\text{small}(\vec{x_i} | \vec{q}_{\true, 7\ldots 260})  & = &  \sum_{a=0}^6 \sum_{b=0}^a \left(1 + a + b\right) ^ {-0.25} \left[ q_{\true, 7+4b} \cos (k_x \, x_i) + q_{\true, 8+4b} \sin (k_x \, x_i) \right] \\
& & \qquad \qquad \times \left[ q_{\true, 9+4b} \cos (k_y \, y_i) + q_{\true, 10+4b} \sin (k_y \, y_i) \right] \quad ,
\end{eqnarray*}
where
\begin{eqnarray*}
k_x & = & \dfrac{a \, \pi}{X} \quad ,\\
k_y & = & \dfrac{b \, \pi}{Y} \quad ,\\
\end{eqnarray*}
with the physical focal-plane dimensions $X$ and $Y$. The $(1 + a +d)^{-0.25}$ factor reduces the magnitude of the higher order terms. In total, the instrument response model is parameterized with 260 parameters; an example can be seen in Figure \ref{fig:single_image} (right). It is this instrument response that we try and recover with the self-calibration procedure. With this instrument response model, we have assumed that the variations in sensitivities on a pixel-to-pixel-scale have been calibrated with some other method, such as an internal flat-field calibration source. It is only the variations on larger scales that we aim to constrain with the self-calibration procedure. Our final assumption in these simulations is that the instrument response is temporally stable. 

\subsection{Complete Survey}
In this work we are interested in simulating complete surveys, in order to identify the crucial characteristics of survey strategies that allow for the relative instrument response model to be accurately, retrospectively constrained from the resulting catalog. We therefore apply the single exposure procedure introduced in the previous section for each pointing specified in a defined survey strategy, which is a set of pointings $(\alpha, \beta)$ and orientations $(\theta)$. The resultant source measurements are collated into a survey-wide catalog. 

\section{Self-Calibrating the Survey-Wide Catalog}
\label{sec:self_cal}
We self-calibrate the catalog generated in the survey simulations to recover an optimum estimate of the true instrument response applied and the true source fluxes. This self-calibration procedure has been successfully applied to ground-based imaging surveys, such as the Sloan Digital Sky Survey \citep{pad08}, where the procedure is dubbed ``uber-calibration'', and the Deep Lens Survey \citep{wit11}. This iterative procedure comprises two steps: (1) a refinement of the source flux estimates based on the latest instrument response model and (2) a refinement of the instrument response model based on the updated source flux estimates. These steps are iterated until the system converges, or until it is clear that the system will not converge. There is a degeneracy in the problem, as both the true instrument response and the true source magnitudes are unknown. It is therefore only possible to calibrate the \textit{relative} instrument response and the \textit{relative} source fluxes. It is not possible to know, for example, if the sources are all fainter or if the instrument response is uniformly lower. In practice, this degeneracy can be broken through observations of a small number of standard absolute sources.

\subsection{Fitting the Instrument Response Model}
To introduce a realistic level of ignorance into the simulations, we use the self-calibration procedure to fit a model that is \textit{incomplete} in two ways. Firstly, the \textit{fitted} instrument response is modeled as an eighth order polynomial, and not the second order polynomial superimposed with sine and cosine contributions used to model the \textit{true} instrument response. Secondly, the assumed measurement uncertainty variances do not include the additional random measurement error $\epsilon_{i}$ introduced in Subsection \ref{sec:noise}. We therefore fit for the following, incomplete measurement model:
\begin{eqnarray*}
c_i = f(\vec{x_i} | \vec{q}) \, s_{k} + e_{i} \quad ,
\end{eqnarray*}
where $c_i$ is the recorded count rate, $f(\vec{x_i} | \vec{q})$ is the fitted instrument response model at a focal-plane position $\vec{x}_i$, $\vec{q}$ is a vector parameterizing the eighth order polynomial instrument response model, $s_k$ is the model source flux estimate and the error $e_i$ is drawn from the Normal Distribution $N(e|0,{\sigma_i}^2)$, such that
\begin{eqnarray*}
\sigma_{{i}}^{2} = \delta^{2} + \eta^{2}\, c^{2}_i \quad ,
\end{eqnarray*} 
where $\delta$ and $\eta$ are the parameters set by the instrument's $10\upsigma$ detection and the S/N $<500$ limits. The $\epsilon_i$ error contribution is intentionally not included in order to simulate systematic problems with the instrument noise model. 

We make one further assumption during this analysis phase, namely that clearly variable sources have been removed from the source catalog, and that the remaining sources vary by less than $\epsilon_{max}$.

\subsection{Step 1: Source Flux Refinement}
The sources are considered individually in the first step of the self-calibration procedure; their flux estimates are refined based on the latest fitted instrument response parameters $\vec{q}$. An error function $\chi^2_{k}$ for all the measurements $i$ of a source $k$ ($i \in \mathcal{O}(k)$) is constructed:
\begin{eqnarray*}
\chi^2_{k} = \sum_{i \in \mathcal{O}(k)} \frac{(c_i-f_{i}(\vec{x_i} | \vec{q}) \, s_{k})^2}{{\sigma_i}^2} \quad ,
\end{eqnarray*}
where $c_i$ are the measured count rates, $f(\vec{x_i} | \vec{q})$ is the fitted instrument response model at a focal-plane position $\vec{x}_i$ and $\sigma_i$ is the assumed noise model. A new estimate of the model source flux $s'_{k}$ is then found by minimizing the error function with respect to the old model source flux $s_{k}$:
\begin{eqnarray*}
\frac{d\chi^2_{k}}{d s_{k}} = \sum_{i \in \mathcal{O}(k)} \frac{-2 f_{i}(\vec{x_i} | \vec{q}) \, (c_i-f_{i}(\vec{x_i} | \vec{q}) \, s'_{k})}{{\sigma_i}^2} = 0 \quad ,
\end{eqnarray*}
\begin{eqnarray*}
s'_{k} \leftarrow \left[{\sum_{i \in \mathcal{O}(k)}  \frac{f_{i}(\vec{x_i} | \vec{q})^2}{{\sigma_i}^2}} \right]^{-1}  \left[ {\sum_{i \in \mathcal{O}(k)} \frac{f_{i}(\vec{x_i} | \vec{q}) \, c_i}{{\sigma_i}^2}} \right] \quad .
\end{eqnarray*}
The standard uncertainty variance on the new source flux estimate $s'_{k}$ is given by
\begin{eqnarray*}
\sigma'^2_k = \left[{\sum_{i \in \mathcal{O}(k)}  \frac{f_{i}(\vec{x_i} | \vec{q})^2}{{\sigma_i}^2}} \right] \quad .
\end{eqnarray*}

\subsection{Step 2: Instrument Response Refinement}
The instrument response parameters can now be refined with the latest source flux estimates. A error function for all the measurements of all the sources is constructed
\begin{eqnarray*}
\chi^2 = \sum_{k} \chi'^2_k \quad ,
\end{eqnarray*}
where
\begin{eqnarray*}
\chi'^2_k = \sum_{i \in \mathcal{O}(k)} \frac{(c_i-f_{i}(\vec{x_i} | \vec{q}) \, s'_{k})^2}{{\sigma_i'}^2}  \quad .
\end{eqnarray*}
Recall that the fitted instrument response $f_{i}(\vec{x_i} | \vec{q})$ is modeled as an eighth order polynomial. This can be expressed as
\begin{eqnarray*}
f_{i}(\vec{x_i} | \vec{q}) = \sum_{l = 1}^L q_{l} \, g_l(\vec{x_i})  \quad ,
\end{eqnarray*}
where $L = 45$ in this case. The total error function $\chi^2$ can be rewritten as
\begin{eqnarray*}
\chi^2 =\sum_{k} \sum_{i \in \mathcal{O}(k)} \frac{(c_i- s'_{k} \sum_{l = 1}^L q_{l} g_l(\vec{x_i}))^2}{{\sigma_i'}^2}   \quad .
\end{eqnarray*}
To refine the instrument response model fit, this error function is minimized with respect to the instrument response model parameters $q_{l}$
\begin{eqnarray*}
\frac{d\chi^2}{dq_{l}} = \sum_{k} \sum_{i \in \mathcal{O}(k)} \frac{-2 g_l(\vec{x_i}) s'_{k} (c_i- s'_{k} \sum_{l' = 1}^{L'} q'_{l'} g_{l'}(\vec{x_i}))}{{\sigma_i'}^2} = 0  \quad ,
\end{eqnarray*}
\begin{eqnarray*}
\sum_{k} \sum_{i \in \mathcal{O}(k)} \frac{g_l(\vec{x_i}) s'_{k} c_i}{{\sigma_i'}^2} = \sum_{k} \sum_{i \in \mathcal{O}(k)} \frac{g_l (\vec{x_i}) s'^2_{k} \sum_{l' = 1}^{L'} q'_{l'} g_{l'} (\vec{x_i})} {{\sigma_i'}^2}   \quad .
\end{eqnarray*}
It is now simpler to proceed in matrix notation. The following substitutions can be made
\begin{eqnarray}
b_l & = & \sum_{k} \sum_{i \in \mathcal{O}(k)} \frac{g_l(\vec{x_i}) s'_{k} c_i}{\sigma_i'^2}    \quad ,\\
G_{ll'} & = & \sum_{k} \sum_{i \in \mathcal{O}(k)} \frac{{s'}_{k}^2}{\sigma_i'^2} g_l(x_i) g_{l'}(x_i)   \quad .
\end{eqnarray}
The matrix equation is then
\begin{eqnarray*}
\vec{b} = G \cdot \vec{q'}    \quad .
\end{eqnarray*}
The refined instrument response parameters are then found by
\begin{eqnarray*}
\vec{q'} \leftarrow G^{-1}  \cdot \vec{b}   \quad .
\end{eqnarray*}
These two steps are iterated until the solution converges to a final fit of the instrument response $f_\fit(\vec{x} | \vec{q_\fit})$ and the source fluxes $s_{k,\fit}$, or until it is clear that a solution will not be found. 

It is worthy of note that although this back-and-forth (Step 1, Step
2) iteration scheme works well and is easy to understand, there is no
reason in principle not to simply put the whole system into a
non-linear optimizer.  A sufficiently clever general-purpose optimizer
might outperform these bilinear iterations.  At the scale of the
simulations and optimizations performed in this paper, there is no
need to look for higher performance algorithms than the bilinear
iterative solution presented here.

\section{Performance Metrics}
\label{sec:metrics}
To assess the performance of the self-calibration procedure with different survey strategies, it is necessary to quantify the quality of the final fitted solution. To do this we defined three quantities. The first is the root-mean-squared (RMS) error $S_\rms$ in the final fitted source fluxes $s_{k,\fit}$ compared to the true source fluxes $s_{k,\true}$ for the $K$ sources used in the self-calibration procedure:
\begin{equation}
S_\rms = \sqrt{\dfrac{1}{K} \sum_k^K \left( \dfrac{s_{k,\fit} - s_{k,\true}}{s_{k,\true}} \right)^2}   \quad .
\end{equation}
In detail, we compute this $S_\rms$ excluding sources near the survey boundary. The other two metrics, called ``badnesses'', are defined as the error between the final fitted instrument response and a reference instrument response sampled on a regular $300 \times 300$ grid across the focal plane. For the ``true badness'' $B_\true$, the \textit{fitted} instrument response $f_\fit(\vec{x} | \vec{q_\fit})$ is compared to the \textit{true} instrument response $f_\true(\vec{x} | \vec{q_\true})$ at the $J$ sample points  
\begin{equation}
B_\true = \sqrt{\dfrac{1}{J} \sum_j^J \left(\dfrac{f_\fit(\vec{x_j} | \vec{q_\fit}) - f_\true(\vec{x_j} | \vec{q_\true})}{f_\true(\vec{x_j} | \vec{q_\true})}\right)^2}   \quad .
\end{equation}
The ``best-in-basis badness'' $B_\basis$ compares the \textit{fitted} instrument response $f_\fit(\vec{x_j} | \vec{q_\fit})$ to the \textit{best instrument response fit possible} $f_\basis(\vec{x_j} | \vec{q_\basis})$ with the basis used to describe the fitted model (in this case an eight order polynomial) at the $J$ sample points
\begin{equation}
B_\basis = \sqrt{\dfrac{1}{J} \sum_j^J \left( \dfrac{f_\fit(\vec{x_j} | \vec{q_\fit}) - f_\basis(\vec{x_j} | \vec{q_\basis})}{f_\basis(\vec{x_j} | \vec{q_\basis})} \right)^2}   \quad .
\end{equation}
The best-in-basis badness is always smaller than the true badness, as it does not include the errors associated with the limitation of the basis used for the fitted model. The badnesses provide a more complete description of the self-calibration performance than the RMS error on the fitted sources' fluxes, as the RMS source error only applies to the bright sources within the survey selected for the self-calibration procedure.

\section{Simple Survey Strategies}
\label{sec:simple_surveys}
In this paper, we consider four simple but very different survey strategies. We use the end-to-end catalog simulation and self-calibration chain, introduced in Sections \ref{sec:survey_simulations} and \ref{sec:self_cal}, to evaluate the performance of the self-calibration procedure with these strategies. The parameters of these simulations are summarized in Table \ref{tab:parameters}.

\subsection{Survey Description}
We label the four survey strategies---which all cover the same 64~deg$^2$ patch of sky---with the letters A to D. These survey strategies are summarized in Table \ref{tab:simple_surveys} and are shown in Figure \ref{fig:simple_surveys}. Strategy A is the simplest strategy; the target field is regularly tiled with small overlaps between adjacent pointings ($\sim 12$~percent in both of the camera pointing directions). The pointings in the 9 passes over the same field are exactly aligned. Survey B is the same as A, but with each pass over the target field the orientation of the telescope is rotated by $40^\circ$. 

Survey C is more complex. The first pass over the field is the same as in Survey A, with $12 \times 12$ pointings. In the next pass, one of the pointings in the $\alpha$ direction is removed and one is added in the $\beta$ and the pointings are respaced uniformly, so the resultant pointing grid is $13 \times 11$. In the third pass over the field this change is reversed and the field is measured on a $11 \times 13$ grid. These three passes are then repeated three times. 

The pointing positions in Survey D are quasi-random: the pointings are the same as with Survey A, but each has a random offset within [-0.375,0.375]~deg applied in both the $\alpha$ and $\beta$ directions. By fixing the pointings within these $0.75~\text{deg} \times 0.75~\text{deg}$ boxes, we ensure that the quasi random strategy has a uniform coverage of the field; that is, the fluctuations in coverage are less than they would be in a Poisson (totally random) process. The orientations of the pointings in Survey D are completely random. 

Surveys A and C can be executed without rotating the camera with respect to the celestial coordinates. Surveys B and D can only be executed with systems that permit rotations of the camera.

\subsection{Self-Calibration Performance}
\begin{figure}
\includegraphics[width=\textwidth]{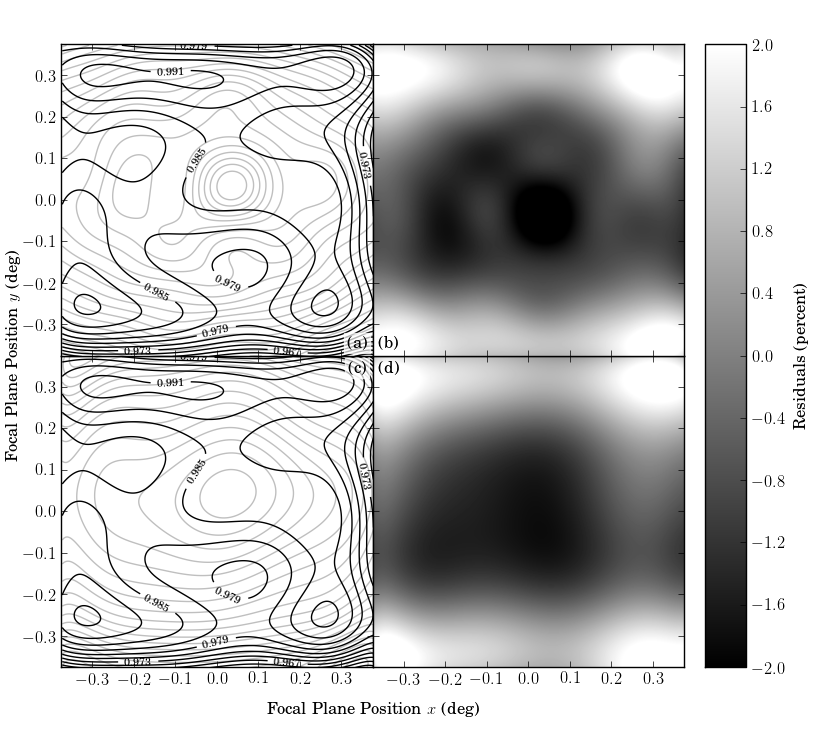}
\figcaption{Survey Strategy A: A comparison of the fitted instrument response model $f_\fit(\vec{x_j} | \vec{q_\fit})$ (black) obtained from self-calibrating the Survey A catalog compared to (a) the true $f_\true(\vec{x} | \vec{q_\true})$ and (c) the best-in-basis $f_\basis(\vec{x_j} | \vec{q_\basis})$ instrument response models (gray). The self-calibration procedure did not converged to the final solution and the results presented here correspond to the instrument response found after 5,000 iterations. The best-in-basis instrument response is the best fit to the true instrument response possible with the basis used to model the instrument response in the self-calibration procedure (in this case an eighth order polynomial). The plots (b) and (d) show the residuals between the two instrument response models plotted in (a) and (c) respectively. With this survey strategy, the center of the focal plane is never connected to other parts with repeat observations of the same sources. \label{fig:A_result}}
\end{figure}

\begin{figure}
\includegraphics[width=\textwidth]{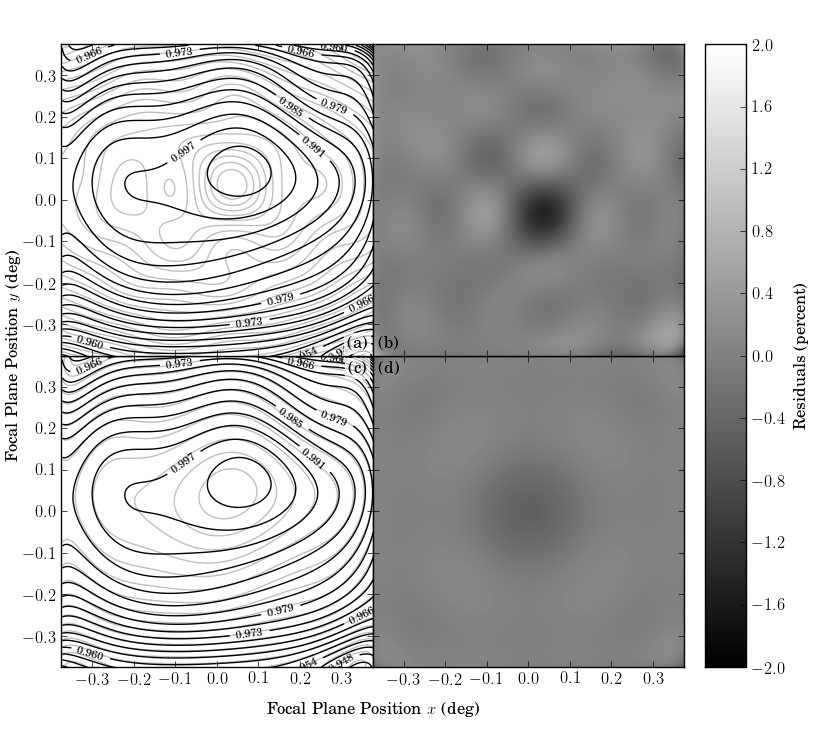}
\figcaption{Survey Strategy B: A comparison of the fitted instrument response model $f_\fit(\vec{x_j} | \vec{q_\fit})$ (black) obtained from self-calibrating the Survey B catalog compared to (a) the true $f_\true(\vec{x} | \vec{q_\true})$ and (c) the best-in-basis $f_\basis(\vec{x_j} | \vec{q_\basis})$ instrument response models (gray). The self-calibration procedure converged to the final solution after 1,114 iterations. The plots (b) and (d) show the residuals between the two instrument response models plotted in (a) and (c) respectively. With this survey strategy, it is only regions towards the edge of the focal plane that are well interconnected with repeat observations of the same sources, and therefore the self-calibration procedure can only fit a reasonable instrument response model at these positions.\label{fig:B_result}}
\end{figure}

\begin{figure}
\includegraphics[width=\textwidth]{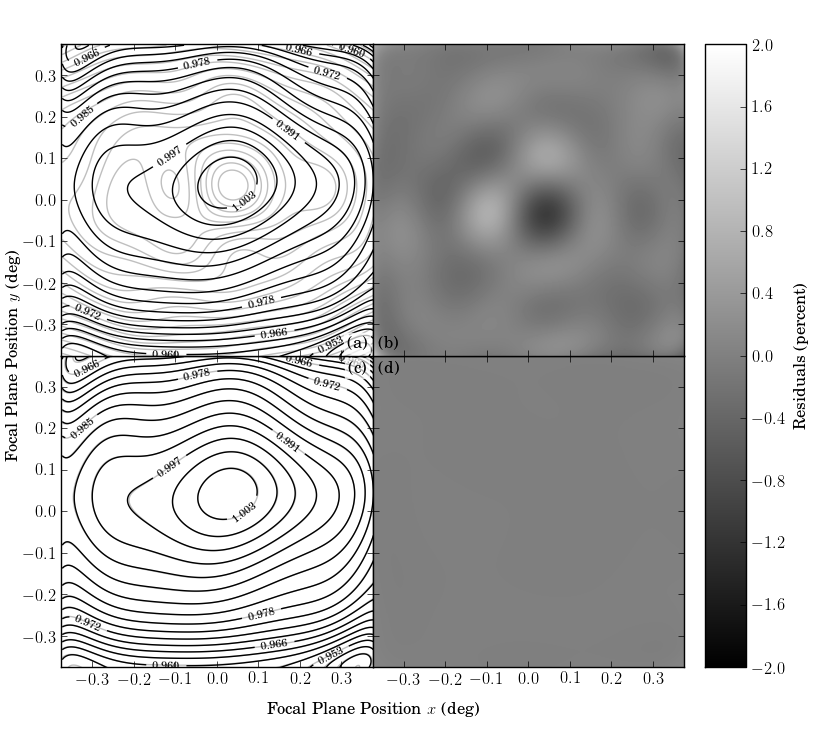}
\figcaption{Survey Strategy C: A comparison of the fitted instrument response model $f_\fit(\vec{x_j} | \vec{q_\fit})$ (black) obtained from self-calibrating the Survey C catalog compared to (a) the true $f_\true(\vec{x} | \vec{q_\true})$ and (c) the best-in-basis $f_\basis(\vec{x_j} | \vec{q_\basis})$ instrument response models (gray). The self-calibration procedure converged to the final solution after 24 iterations. The plots (b) and (d) show the residuals between the two instrument response models plotted in (a) and (c) respectively. With this regular observing strategies, all regions of the focal plane well interconnected with each other.\label{fig:C_result}}
\end{figure}

\begin{figure}
\includegraphics[width=\textwidth]{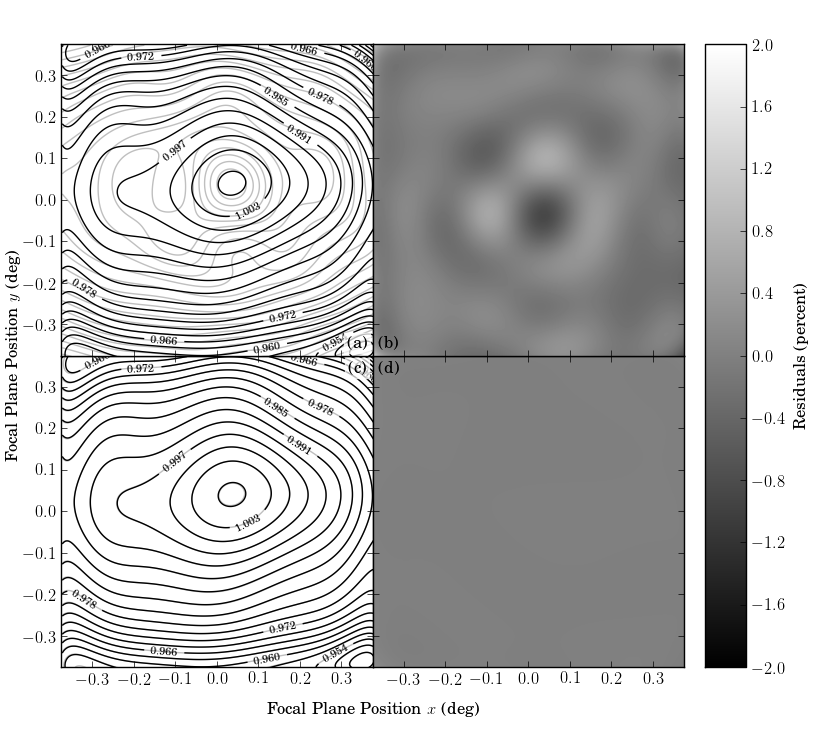}
\figcaption{Survey Strategy D: A comparison of the fitted instrument response model $f_\fit(\vec{x_j} | \vec{q_\fit})$ (black) obtained from self-calibrating the Survey D catalog compared to (a) the true $f_\true(\vec{x} | \vec{q_\true})$ and (b) the best-in-basis $f_\basis(\vec{x_j} | \vec{q_\basis})$ instrument response models (gray). The self-calibration procedure converged to the final solution after 11 iterations. The plots (b) and (d) show the residuals between the two instrument response models plotted in (a) and (c) respectively. With this quasi-random observing strategies, all regions of the focal plane well interconnected with each other.\label{fig:D_result}}
\end{figure}

The iterative self-calibration procedure converges to a final fitted solution for the Survey B, C and D catalogs. With the Survey A catalog it does not converge in any reasonable time period, even when the system is started close to the optimum fit. The fitted instrument response solutions from the Surveys A to D catalogs are shown in Figures \ref{fig:A_result} to \ref{fig:D_result}. In these plots the final fitted instrument response $f_\fit(\vec{x_j} | \vec{q_\fit})$ is compared to the true instrument response $f_\true(\vec{x} | \vec{q_\true})$ and to the best possible instrument response fit with the basis used $f_\basis(\vec{x_j} | \vec{q_\basis})$ (``best-in-basis''). The accuracy of the final fitted solutions are summarized, with the metrics introduced in Section \ref{sec:metrics}, for each of the Survey strategies in Table \ref{tab:simple_results}. 

The survey strategies, the resultant sky coverages and the errors in final self-calibration flux solutions are summarized for each survey individually in Figures \ref{fig:A_survey} to \ref{fig:D_survey}. With Surveys A and B---and to a lesser extent Survey C---the imprint of the strategies can be clearly seen in the coverage maps. It is important to note that---with sufficient re-visits to the field---the coverage of the quasi-random survey D is more uniform than the other strategies investigated.

The self-calibration procedure performs best with the Survey C and Survey D catalogs. The performance of the self-calibration procedure is worse for the Survey B catalog and it fails completely with the Survey A catalog, where no solution was obtained. The number of iterations of the self-calibration procedure required to converge to the final solution also decreases as the quality of the final fit increases: Survey D required 11 iterations, Survey C 24, Survey B 1,114 and the procedure was iterated 5,000 times with Survey A without reaching a solution.

With the Survey C and D catalogs, the instrument response is accurately recovered. As can be seen from Figures \ref{fig:C_result} and \ref{fig:D_result} (top), the fitted instrument response $f_\fit(\vec{x_j} | \vec{q_\fit})$ does not perfectly reproduce the true instrument response $f_\true(\vec{x} | \vec{q_\true})$. That said, through comparisons with the best fit possible with the eight order polynomial basis used in the self-calibration procedure $f_\basis(\vec{x_j} | \vec{q_\basis})$ (bottom), we find that majority of the remaining errors in our fits are due to limitations in the model basis used. The residuals in this comparison are low; this can also be clearly seen in the significant improvement in the best-in-basis badness $B_\basis$ compared to the true badness $B_\true$. The results from these two survey catalogs are clear: if the correct redundancy is built into the survey, then self-calibration calibration can be used to accurately constrain the relative instrument response, with presumably the majority of the remaining errors coming from the limitations in the model used to represent the instrument response. 

Surveys C and D perform comparably well, although they are very different in their configuration. Two substantial advantages of Survey C over Survey D are: In Survey C the camera would not need to be rotated relative to the sky between pointings; indeed many survey imaging cameras have fixed celestial orientations. Survey D requires generating quasi-random numbers, whereas Survey C has a completely deterministic set of pointing centers. This is a conceptual advantage to Survey C, but possibly also an operational advantage. 

The Survey A and B catalogs also give interesting results. With Survey A, the self-calibration procedure did not reach a solution. This---in itself---is an important result, as the naive tiling of the sky currently being considered for many large surveys will result in catalogs that cannot be retrospectively self-calibrated. Minor changes to survey strategies, such as those in Survey C, can drastically improve the performance of the self-calibration procedure. Without these changes, vast amounts of calibration information is simply lost. We can also draw conclusions from the results obtained with Survey B: here the self-calibration procedure converges close to the correct solutions, albeit not as close as the Surveys C and D. Here, it is the quality of the fit over the focal plane that is interesting. From the residual plot against the best-in-basis instrument response $f_\basis(\vec{x_j} | \vec{q_\basis})$ (Figure \ref{fig:B_result}d), we can see that the quality of the fit is rotationally symmetric, with the self-calibration procedure able to fit the instrument response well at the edge of the focal plane, but poorly at the center. As a result, the fluxes of the sources that fall at the center of the focal plane during a measurement cannot be constrained as well as those that fall at the edges; this can clearly be seen in the post-self-calibration flux error map shown in Figure \ref{fig:B_survey}d.

This, along with the success of the procedure with the Surveys C and D catalogs, gives an insight into the key property of a survey that makes it good for self-calibration. Namely, the survey should return the same sources to many, very different focal-plane positions. This allows the self-calibration procedure to link the different parts of the focal plane with each other through the observations of the same sources. Surveys C and D do this, and the self-calibration procedure performs well with their catalogs. Survey A does not do this; it only connects the outer edges of the focal plane to each other. There are no redundant measurements connecting the majority of the focal plane to other positions. This conclusion in confirmed with Strategy B. Here, positions at the outside of the focal plane are well connected to other positions, and therefore the self-calibration procedure can well fit the instrument response in these regions. In the center of the focal plane, on which the same sources always fall, there are insufficient redundant measurements to allow for a good fit. Returning the same sources to very different focal-plane positions is critical for the self-calibration to work effectively. 

\section{Discussion}
From the simulations presented in this paper, we are able to draw two firm conclusions about the self-calibration of large-scale imaging surveys. The first is that, in agreement with previous work, we show that this technique is very powerful, allowing the large-scale, relative instrument response of an imaging instrument to be accurately retrospectively constrained from the science images alone. The second is that there must be the correct redundancy built into the survey strategy. The same sources must be imaged at very different focal-plane positions; regular tiling of the sky is useless for this kind of calibration. 

As this work was completed in the context of space-based imaging surveys, we have not considered the temporal variation of the relative instrument response as a high priority. We concede that this will have a significant impact on ground-based surveys, where variations on nightly timescales can be expected. The self-calibration procedure discussed in this paper can be trivially modified to fit for a time evolving relative instrument response model. For this to be effective though, it will be necessary to not only observe the same sources at very different focal-plane positions, but also to observe them at very different times within the survey.  

In this paper, we have considered the design of survey strategies for calibrating the large-scale, relative photometric response of an instrument, but there are many more calibration parameters that can be constrained from this method. For example, the optical distortion of an instrument or the point-spread-function variation could also be constrained with such a method, although it would need to be explored whether the properties of a survey strategy that makes it good for one calibration, are also suitable for other calibration aspects.

Ultimately, survey design requires trade-offs between calibration requirements and other requirements related to cadence, telescope overheads, and observing constraints (like alt, az, airmass, and field-rotation constraints).  One requirement that is often over-emphasized, however, is conceptual or apparent ``uniformity''; these different survey strategies have different uniformities of coverage, each with possibly non-obvious consequences.  Many astronomers will see survey strategy A as being ``more uniform'' in its coverage (especially relative to strategy D).  This is not true, if uniformity is defined by the variance or skewness or any other simple statistic in exposure times; strategy D is extremely uniform (more
uniform than Poisson, for example).  In any survey, past and future, variations in exposure time have been valuable for checking systematic and random errors, and don't---in themselves---make it difficult to obtain a uniform survey in properties like brightness (since samples can be cut on any well-measured property).  In general, in the presence of real astronomical variations in distributions of luminosity, distance, metallicity, and (more importantly) extinction by dust, there is no way to make a survey uniformly sensitive to the objects of interest.  As a community we should be prepared to adjust our analysis for the non-uniformity of the survey rather than adjust (cut) our data to match the uniformity of unrealistically simplified analyses.  This is already standard practice in the precise cosmological measurement experiments, and will be required practice for the next generation of massively-multiple-epoch imaging surveys.

\acknowledgments
It is a pleasure to thank Rob Fergus (NYU), Jason Kalirai (STScI), Sam Roweis (deceased), and Pieter van Dokkum (Yale) for valuable conversations. 

Rory Holmes was funded by the Deutschen Zentrums f\"ur Luft- und Raumfahrt
(DLR) through the grant 50QE1202.  David W. Hogg was partially supported by NASA
(grant NNX08AJ48G), the NSF (grants AST-0908357 and IIS-1124794), and
a research fellowship from the Alexander von Humboldt Foundation.

The code used in this work is publicly available at \url{github.com/davidwhogg/SelfCalibration/}~.

\clearpage

\clearpage
\begin{deluxetable}{lc}
\tablewidth{0pt}
\tablecaption{A summary of the tunable parameters in the simulations and their fiducial values.\label{tab:parameters}}
\tablehead{
\colhead{Parameter} & \colhead{Fiducial Value}}
\startdata
Source Density -- Eqn. \ref{eqn:power_law} (deg$^{-2}$) & $a=-13.05, b=1.25, c=-0.02$ \\
Survey Area (deg$^2$)& $8 \times 8$ \\
Source Density (deg$^{-2}$) & $ d = 300 $ \\
Saturation Limit (mag) & $m_\text{min} = 17$\\
10$\upsigma$ Detection Limit (mag)& $m_\text{max} = 22$\\
Field-of-View (deg$^{2}$)& $0.75 \times 0.75$ \\
Noise Model -- Eqn. \ref{eqn:noise} & $\delta = 0.1585$, $\eta = 0.0017$, $\epsilon_\text{max} = 1.0$\\
Fitted Instrument Response Model & 8$^{\text{th}}$ order polynomial\\
\enddata
\end{deluxetable}

\clearpage
\begin{deluxetable}{cccc}
\rotate
\tablecaption{A summary of the four simple survey strategies considered in this paper. \label{tab:simple_surveys}}
\tablewidth{0pt}
\tablehead{
\colhead{Survey Label} & \colhead{Pointing Center} & \colhead{Orientation} & \colhead{Number of Pointings}}
\startdata
A  & Uniform Grid ($12\times12$) & $0^\circ$ & 1296 \\
B      & Uniform Grid ($12\times12$)     & Each Pass: $\theta +40^\circ$& 1296 \\
C   & Pass 1:  Uniform Grid ($12\times12$)  &  & 1290\\
 & Pass 2:  Uniform Grid ($13§\times11$) & $0^\circ$&\\
  & Pass 3:  Uniform Grid ($11§\times13$) & &\\
D   & Quasi-Random  & Random & 1296\\
\enddata
\end{deluxetable}

\begin{figure}
\includegraphics[width=\textwidth]{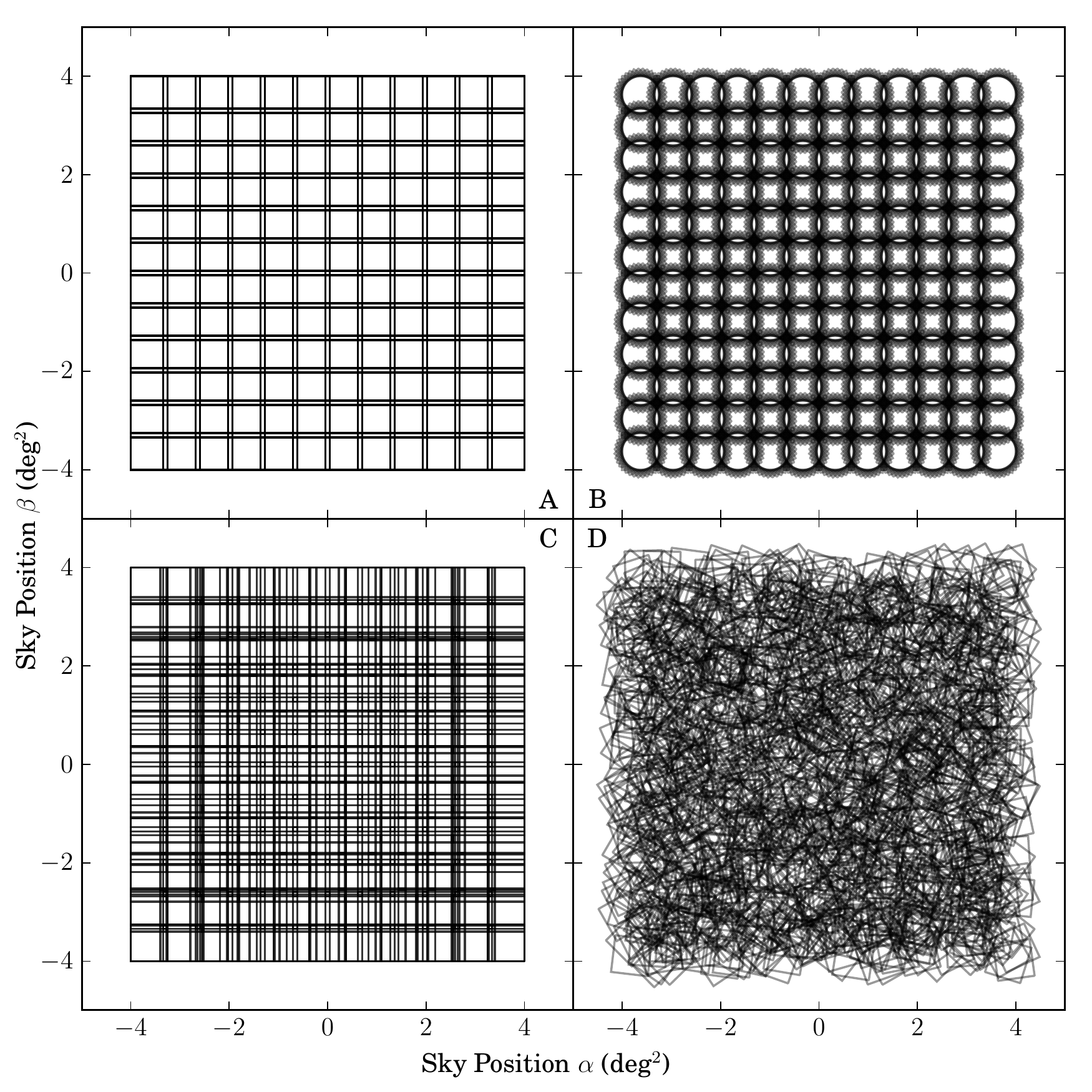}
\figcaption{Focal-plane footprints projected onto the synthetic sky according to the four simple survey strategies described in Section \ref{sec:simple_surveys} and summarized in Table \ref{tab:simple_surveys}. Surveys A, B and D have 1296 pointings and survey C has 1290 pointings.\label{fig:simple_surveys}}
\end{figure}

\newcommand{\surveyfigcaption}{and the self-calibration results: (a) shows the survey footprint diagram (as shown in Figure \ref{fig:simple_surveys}), (b) shows the source coverage for this survey, with (c) showing a histogram of the number of source observations (excluding sources very near the survey boundary) and (d) shows the flux errors for each of the sources after the self-calibration procedure.  In panels (b) and (d) the white gaps are \emph{not} gaps in the survey coverage but rather sky positions at which there do not happen to be observed stars}

\begin{figure}
\includegraphics[width=\textwidth]{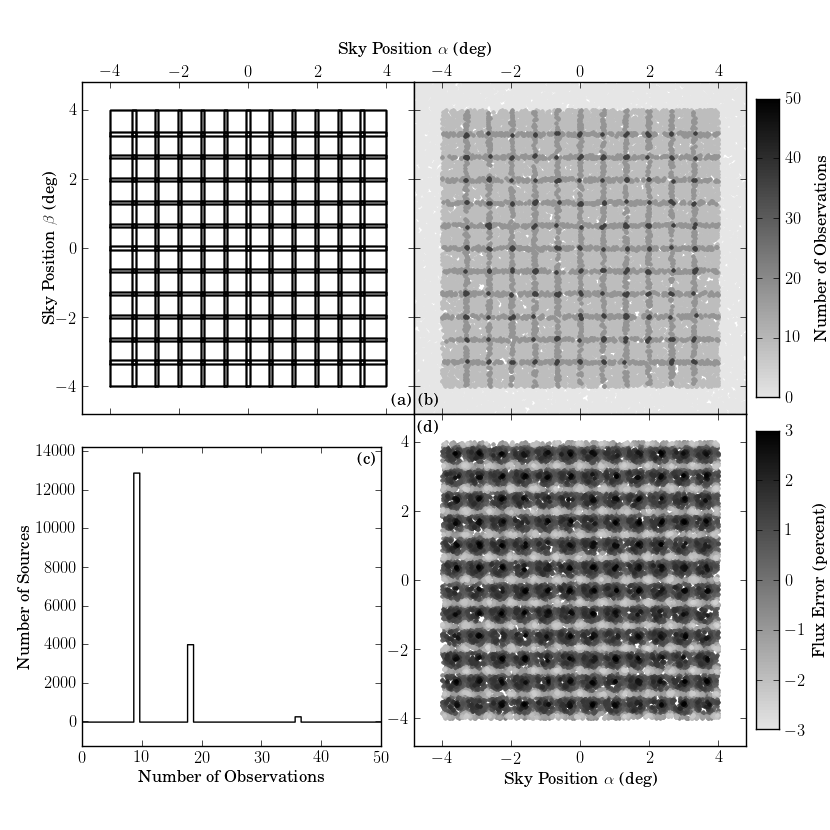}
\figcaption{A summary of Survey Strategy A \surveyfigcaption. The imprint of the survey strategy can be clearly seen in the coverage map. With this survey strategy, the self-calibration procedure is not able to converge to a solution and therefore there are strong and regular residuals in the flux error map.
\label{fig:A_survey}}
\end{figure}

\begin{figure}
\includegraphics[width=\textwidth]{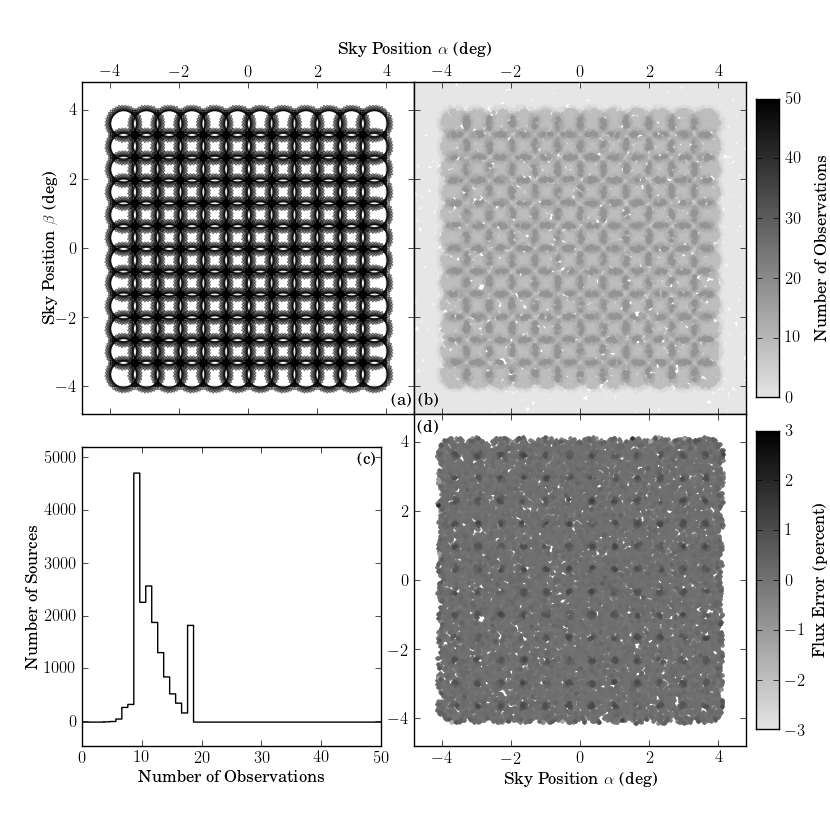}
\figcaption{A summary of Survey Strategy B \surveyfigcaption. The imprint of the survey strategy can be clearly seen in the coverage map. With this survey strategy, the self-calibration procedure can well calibrate the instrument response at the edge of the focal plane, but not in the center. As a result, the residuals in the flux error map are higher at the center of each pointing, where the sources fall on a less well constrained part of the focal plane. 
\label{fig:B_survey}}
\end{figure}

\begin{figure}
\includegraphics[width=\textwidth]{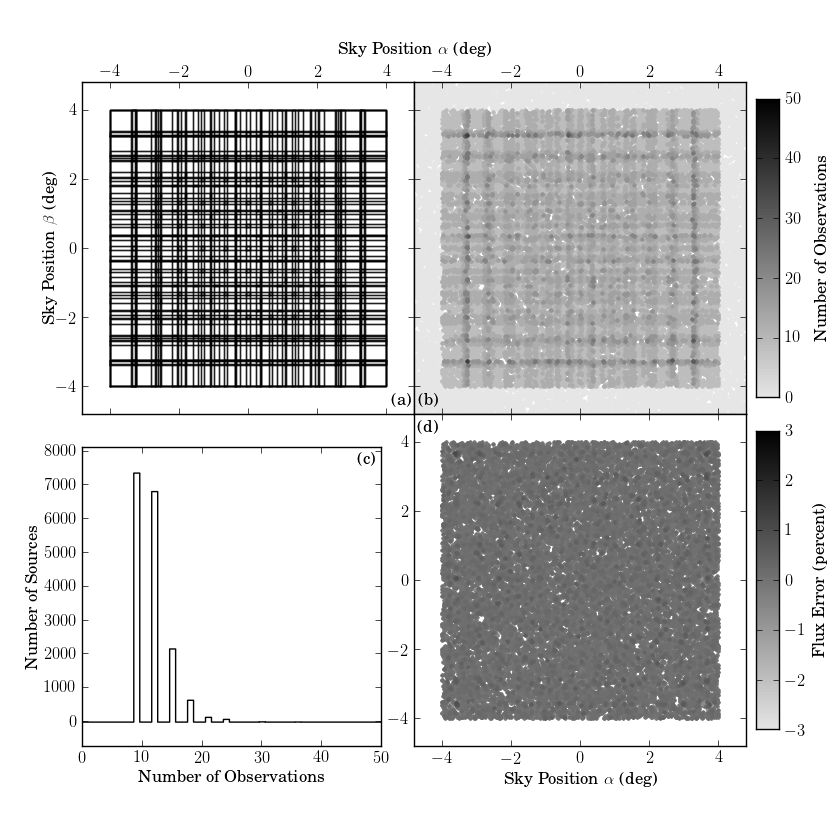}
\figcaption{A summary of Survey Strategy C \surveyfigcaption.
\label{fig:C_survey}}
\end{figure}

\begin{figure}
\includegraphics[width=\textwidth]{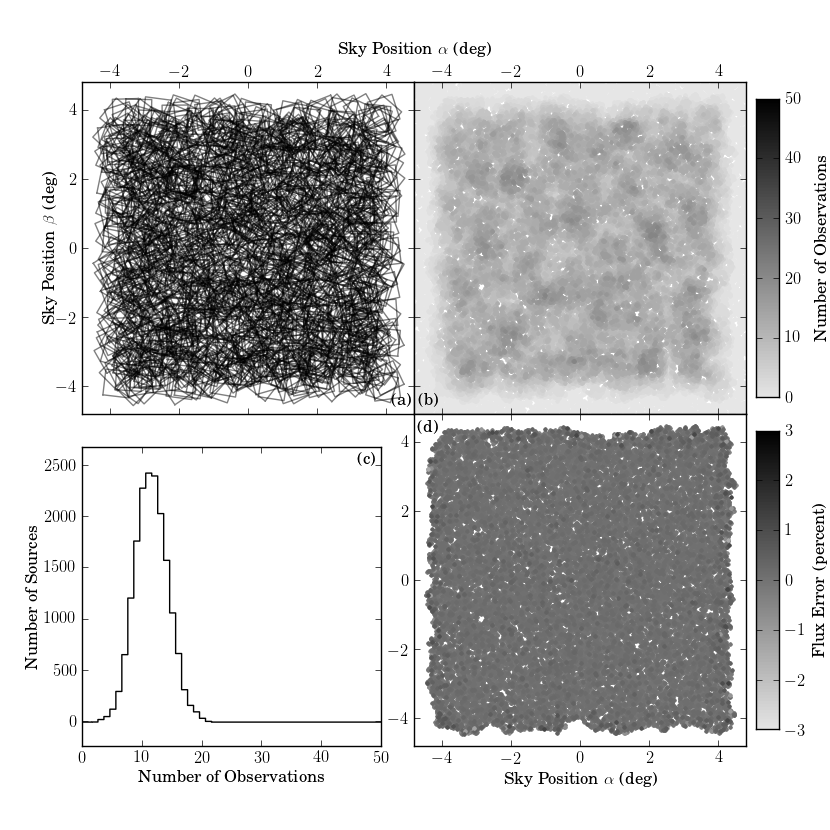}
\figcaption{A summary of Survey Strategy D \surveyfigcaption. As can be seen in panel (c), this quasi-random survey offers a more uniform coverage of the field than the other surveys; it also has much more uniformity than would a Poisson distribution of field centers.
\label{fig:D_survey}}
\end{figure}

\clearpage
\begin{deluxetable}{ccccc}
\rotate
\tablecaption{A summary of the quality of the final fit from the self-calibration procedure with the four simple survey strategies summarized in Table \ref{tab:simple_surveys}. The self-calibration procedure was run with only the brightest sources within the survey area; the source error $S_\rms$ corresponds to the measurements of these sources only and not all those within the survey. *Did not converge.\label{tab:simple_results}}
\tablewidth{0pt}
\tablehead{
\colhead{Survey Name} & \colhead{Iterations} & \colhead{Source Error}       & \colhead{True Badness}        & \colhead{Best-in-Basis} \\
                      &                      & \colhead{$S_\rms$ (percent)} & \colhead{$B_\true$ (percent)} & \colhead{$B_\basis$ (percent)}}
\startdata
A*  & - & 1.172 & 1.680 & 1.527 \\
B  & 1140 & 0.201 & 0.507 & 0.207 \\
C  & 24 & 0.164 & 0.361 & 0.019 \\
D  & 11 & 0.130 & 0.321 & 0.012 \\
\enddata
\end{deluxetable}

\end{document}